\documentclass[aip,rsi,reprint,amsmath,amssymb,
]{revtex4-1}
\pdfoutput=1
\usepackage{graphicx}
\usepackage{epstopdf}
\usepackage{dcolumn}
\usepackage{bm}
\usepackage{amsmath}
\usepackage{amssymb}
\usepackage{gensymb}
\usepackage{textcomp}
\usepackage{multirow}

\begin{document}

\title{Microscopic Characterisation of  Photo Detectors from CMS Hadron Calorimeter}
   



\author{R A Shukla}
\affiliation{Tata Institute of Fundamental Research, Mumbai, India}
\author{V G Achanta}
\affiliation{Tata Institute of Fundamental Research, Mumbai, India}
\author{P D Barbaro}
\affiliation{University of Rochester, New York, USA}
\author{S R Dugad}
\email{dugad@cern.ch}
\affiliation{Tata Institute of Fundamental Research, Mumbai, India}
\author{A Heering}
\affiliation{University of Notre Dame, Indiana, USA}
\author{S K Gupta}
\affiliation{Tata Institute of Fundamental Research, Mumbai, India}
\author{I Mirza}
\affiliation{Tata Institute of Fundamental Research, Mumbai, India}
\author{S S Prabhu}
\affiliation{Tata Institute of Fundamental Research, Mumbai, India}
\author{P Rumerio}
\affiliation{University of Alabama, Tuscaloosa, Alabama, USA }
\affiliation{CERN, Geneva, Switzerland}
%

%

\date{\today}

\begin{abstract}
The CMS hadron Calorimeter is made of alternating layers of scintillating tiles and 
metals, such as brass or iron. The original photo detectors were hybrid units with a 
single accelerating gap called Hybrid Photo Diodes (HPD). Scintillating light was 
transmitted to the HPD's by means of optical fibers. During data  taking at the Large 
Hadron Collider (LHC), the signal strength of scintillator tiles of detector units in 
the forward region degraded significantly due to the damage related to the amount 
of radiation to which the scintillator was exposed to. Scintillators suffer damage when 
exposed to radiation, however, the amount of damage observed was more than 
originally estimated. Several HPDs were removed during a detector shut down 
period. Microscopic scans of relative quantum efficiencies for few of these HPDs 
were made. The damage of the photocathode was determined to vary with the 
amount of optical signal transmitted by optical fibers to the HPD. Imprints of each 
fiber ($\sim$ 1 mm) on the photocathode with varying damage within the same 
pixel were observed. Most of the observed reduction of the calorimeter signal 
can be attributed to localised damage of the photocathode.
\end{abstract}

\maketitle


\section{Introduction}\label{sec_intro}
\begin{figure*}[hbtp]\centering
 \includegraphics{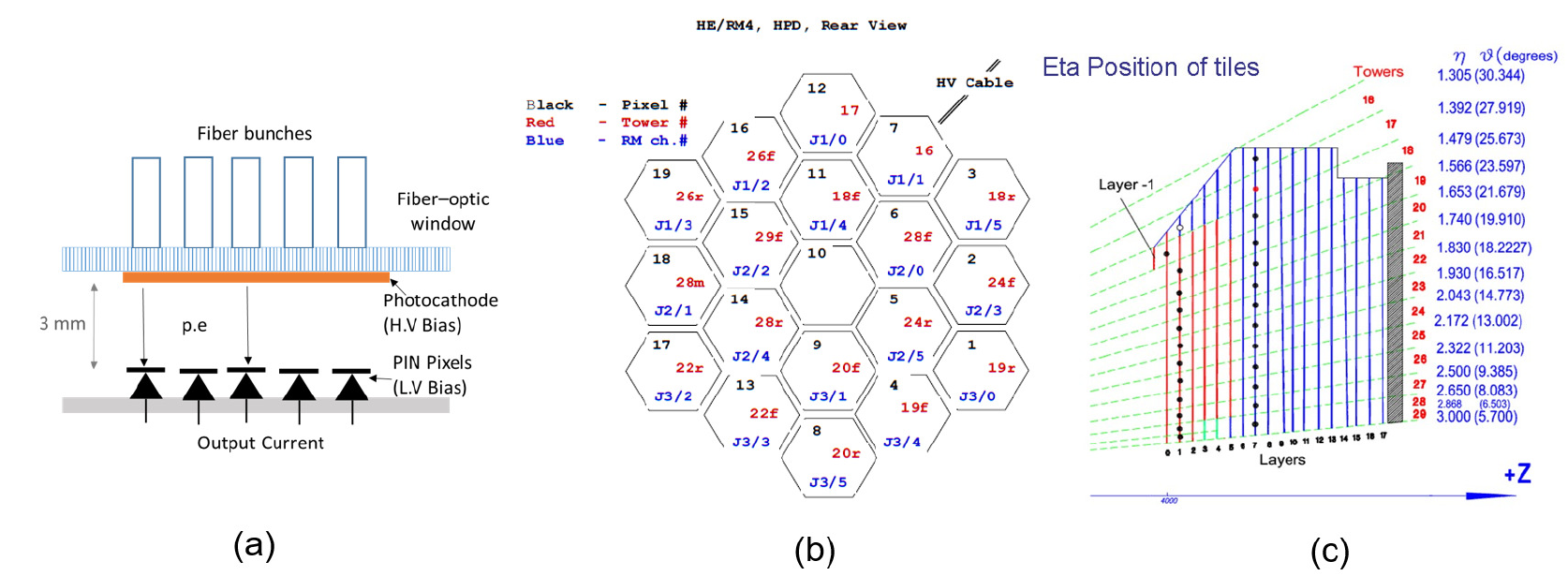}
 \caption{ a) Schematic of the design of HPD;
           b) Typical mapping of HE towers onto an HPD. The number on each 
              pixel represents the pixel number within the HPD, the geometrical 
              tower number and the electronics index on the readout card; and 
           c) Geometrical schematic of the towers in the HE detector.}
 \label{fig:HPD1}
 \end{figure*}

The CMS Hadron Calorimeter~\cite{CMS:1997tfa} is designed to accurately measure 
the energy of jets produced in proton-proton (pp) collisions. The calorimeter consists 
of different sub-detectors, a) Barrel (HB~\cite{Baiatian:1049915,Abdullin2008}), 
b) Endcap (HE~\cite{Baiatian:1103003}),  c) Outer Barrel 
(HO~\cite{Abdullin2008HO, Acharya:973131}) and d) Forward 
(HF~\cite{Abdullin2008B}). The HB, HE, and HO sub-detectors are sampling 
calorimeters and use scintillator embedded with wavelength shifting fibers (WLS) as 
an active element. In the HE and HB detectors, there are several layers of scintillators 
interleaved with an absorber. The detector is segmented in fine $\eta-\phi$ towers, 
where $\eta$ represents the pseudorapidity of a tower, defined as, 

\begin{equation}\label{eqn:eta}
   \eta = -ln(tan(\theta/2)) 
\end{equation}

where $\theta$ is the polar angle of a tower {\it w.r.t.} the pp collision axis and $\phi$ 
represents the azimuth angle of tower in the detector.  The size of tower defined in 
$\eta-\phi$ direction is typically about 0.087$\times$0.087 radians. 
Wavelength-shifting (WLS) fibers spliced to a clear optical fiber are used to transport 
the scintillation light to  the photo-readout elements~\cite{KRYSHKIN1986583}. 
The optical fibers from several sampling layers within the same $\eta-\phi$  tower 
are grouped together inside an {\it Optical Decoding Unit} (ODU) and are mapped 
onto a designated pixel on the HPD. Another fiber, also coupled to the HPD,
brings light produced by an LED, for calibration purposes. Customised Hybrid 
Photo Diodes (HPDs) manufactured by 
DEP~\footnote{DEP (Delft Electronic Products), Dwazziewegen 2\\ Roden,9301 
Netherlands} were installed as photo-readout elements~\cite{ELIAS1997104, 
Cushman:2000iv, CUSHMAN1997107} for these detectors. They have high gain, 
immunity to magnetic field when aligned to HPD axis~\cite{CUSHMAN1998300} 
and a compact size. Each HPD has a hexagonal shape with common photocathode 
and 19 isolated photodiodes underneath, referred to as pixels (p-i-n pixels).  
The HPDs were designed to operate for 10 years in the CMS experiment, 
corresponding to an integrated charge of 3 C/pixel at the highest pseudo-rapidity 
locations~\cite{Cushman:1103002}.  The HPDs have a hexagonal shape.  
Figure~\ref{fig:HPD1}(a) shows a schematic representation of an HPD. The 
geometry of the HE detector is shown in Fig.~\ref{fig:HPD1}(c). As can be seen, 
there are several layers of scintillators in each $\eta-\phi$ tower. Towers with 
higher $\eta$-index are closer to beam pipe, and scintillator layers at shallower 
depth (lower layer number) in a given tower are closer to the pp collision point. 
Hence, layers with higher $\eta$ index and lower depth are expected to produce 
higher scintillation light. The mapping of HE towers on to the pixels of an HPD 
is shown in Fig.~\ref{fig:HPD1}(b). \\

Gradual degradation in the light output of HE detector with increasing 
integrated luminosity  has been observed since its operation. Some 
degradation is expected, due to the presence of very high radiation in this 
region, up to $\approx$ 0.2 Mrad during LHC Run 1, which may cause 
damage to either the scintillator or the HPDs or both. In-situ studies of 
the radiation damage to this detector are carried out using LED, 
radioactive source calibration, and collision data itself. Though these 
studies have shown significant degradation in the overall performance 
of the detector, they do not have the ability to decouple radiation damage 
to the scintillators from damage to the HPDs due to integrated use.
Recently, some HPDs from the HE and HO detectors were removed,
to carry out an independent study of the cumulative effect on them. Since, 
HPDs are primarily exposed to green light from the scintillator, the response 
of the decommissioned HPDs to green light ($\lambda$=520 nm) was 
studied using the {\it Micron Resolution Optical Scanner} (MROS), which was 
specially designed and built for microscopic characterisation of photo 
detectors such as silicon photo-multipliers~\cite{RShukla}. The MROS has 
been demonstrated to provide a focused beam of light on a target. The 
detector under test in this setup can be moved in an automated manner in 
three orthogonal directions with a step size of 0.1 $\mu$m. Using this setup, 
an extensive microscopic characterisation of the HPDs obtained from the HO 
and HE detectors  has been carried out. These studies have revealed 
significant localised damage to the photocathode of the HPD decommissioned 
from the HE detector, whereas the damage observed for the HO HPD is 
quite small and restricted to light incident by the calibration fiber on 
photocathode. Details of the experimental methods used and results that have 
been obtained are discussed in the following sections.

\section{Experimental Setup}

\begin{figure}[hbtp]\centering
 \includegraphics{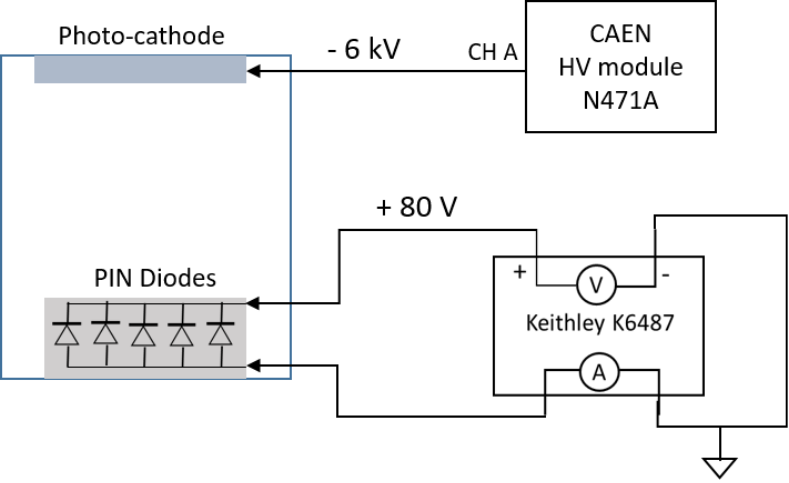}
 \caption{Electrical connections of the HPD module with Bias Voltage for 
          PIN diodes and High Voltage for the photocathode.}
 \label{fig:HPD_eCon}
\end{figure}

\begin{figure}[hbtp]\centering
 \includegraphics{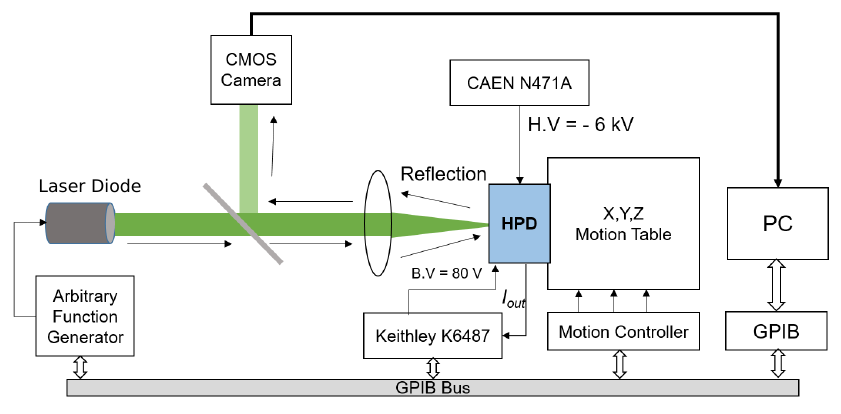}
 \caption{Experimental setup for microscopic characterisation of an HPD.}
 \label{fig:ExpSetup}
\end{figure}

The design of the MROS~\cite{RShukla} is quite suitable for a fine scan of HPD.  
The device under test (HPD) is mounted on a motion table consisting of three 
linear stages capable of motion in three orthogonal directions with a resolution 
of 0.1 $\mu m$ and with a dynamic range of 25 mm. 
Size of the common photocathode of an HPD is also about 25 mm, thus, almost
the entire surface area of the device can be scanned. The recorded localised 
photo response at each scanning position yields complete map of the device 
response across its photocathode. For ease of 
measurement and integration, the HPD mounting board was modified to 
provide the sum of all PIN diode currents (rather than 19 individual currents)  
into the ammeter. The detailed electrical connection diagram is shown in 
Fig.~\ref{fig:HPD_eCon}. Since the MROS would illuminate an ultra-fine 
spot on a  pixel selectively, while other pixels are in the dark, the net output 
current can still be attributed to a scanned position on the particular pixel.
The MROS also has a built-in imaging capability, which helps in carrying out  a
visual inspection of  the surface of the detector, as well as in selecting a particular 
region of interest to be scanned. It is to be noted that the laser intensity is 
kept low during the scan to avoid any possible damage of the 
photocathode during long runs. The overall experimental setup is shown in 
Fig.~\ref{fig:ExpSetup}. \\

\begin{figure}[hbtp]\centering
 \includegraphics{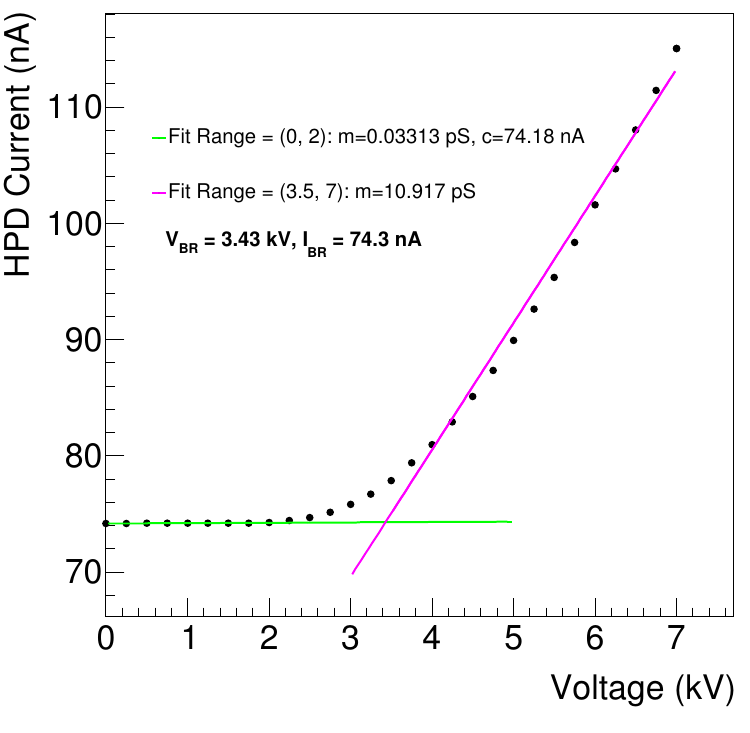}
 \caption{I-V characteristic of the HPD photocathode under constant 
          illumination and a constant photodiode bias of 80 Volts.}
 \label{fig:HPD-HV-IV}
\end{figure}

An HPD requires two power supplies for its operation. As shown in 
Fig.~\ref{fig:HPD_eCon}, a high voltage (negative) bias is applied to the 
photocathode to accelerate the photoexcited electrons emitted by the 
photocathode (photon energy $E_{\gamma} \ge$ work function of 
photocathode). All photo-diodes are connected in parallel for this study. A 
common low voltage (80 Volts) is applied across all the diodes to adequately 
reverse bias them. As can be seen from Fig.~\ref{fig:ExpSetup}, photo-diodes 
are placed at the anode region of the vacuum tube. The high energy electrons 
impinging on the photo-diode give rise to a large number of carriers which are 
further multiplied and efficiently transported (less scattering and recombination) 
to the output electrode due to the presence of the diode bias (electric field). 
Thus, such a combined arrangement yields a large gain of about 2000 with low 
noise. The high cathode voltage at which the energy of the accelerated electrons 
become sufficient to generate detectable current inside photo-diode is referred to 
as the breakdown voltage of the HPD; beyond this voltage the output current 
increases linearly with over voltage (difference between applied and 
breakdown voltage). Typical I-V characteristics of the photocathode (illuminated 
with laser light) with constant photo-diode bias voltage (BV = 80 Volts) is shown 
in Fig.~\ref{fig:HPD-HV-IV}. The breakdown voltage plays an important role in the 
overall operation of the HPD and is discussed in detail in 
Section~\ref{sec:breakdown}. 

\section{Imaging and Focal Axis Determination}
\begin{figure}[hbtp]\centering
  \includegraphics{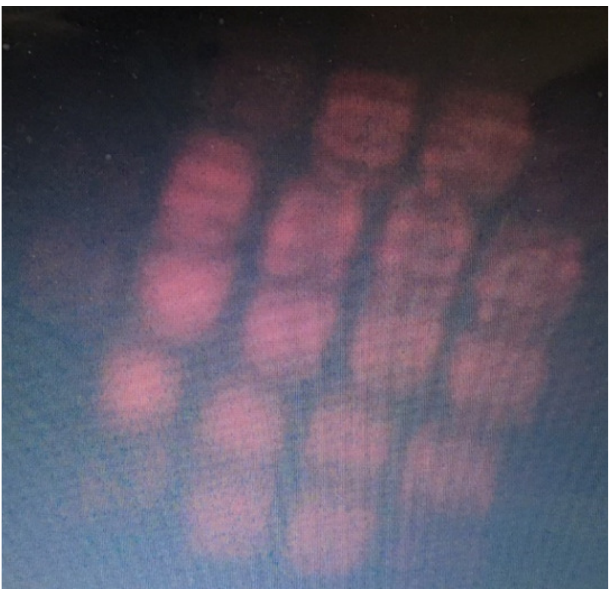}
     \caption{Image of the HPD surface recorded with the built-in CMOS camera.}
     \label{fig:knife-edge_CCD}
\end{figure}

Before an HPD can be microscopically scanned, a built-in CMOS camera system 
was used to survey the surface of the device and access the scan area. The 
imaging also helps 
in roughly determining the focal plane position i.e aligning HPD surface to the focal 
plane. Figure~\ref{fig:knife-edge_CCD} shows a typical image of an HPD surface 
taken with the built-in CMOS camera. The square array pattern is a typical 
characteristic of the fiber optic plate (FOP) of the HPD. The FOP  is an optical 
device comprised of a bundle of micron-diameter optical fibers. The FOP directly 
conveys light incident on its input surface to its output surface. A modified 
knife-edge method was adapted to accurately determine the position of the focal 
plane and to establish the laser beam profile on the photocathode with the fiber 
optic window.

\subsection{Knife-edge Measurement}
In a typical knife-edge method~\cite{RShukla,firester1977}, a sharp-edged object 
is placed in the light beam path between a focusing lens and a detector (e.g a 
PIN diode). The knife-edge is moved perpendicular to the beam direction in fine 
steps, cutting it across and simultaneously recording the PIN diode current to record 
the integrated light falling on its surface at each step. The knife 
edge eventually completely blocks the beam, and the PIN output falls to the typical 
dark current level. The recorded PIN diode current as a function of the vertical 
position of the knife-edge represents an integrated light intensity at each point, and 
hence, when differentiated, yields the actual beam profile (typically a Gaussian). 
The width of the profile (sigma) represents the beam spot size at a particular 
position along the beam axis. A similar exercise can be carried out to obtain the 
beam profile for different positions along the light beam line for the knife edge;  
the position with the smallest beam spot size represents the focal plane of the 
system. The  standard knife-edge method was modified to perform the beam 
profiling without introducing a dedicated knife edge, using instead the HPD 
PIN diode edges, which act as a boundary between the HPD's dead and active 
regions to cut the beam. When the laser beam is completely in the dead area 
(outside PIN diode boundary), the observed HPD current is equal to its typical 
dark current. This situation is equivalent to completely blocking the optical beam. 
As the beam is scanned across the HPD, and the beam slowly approaches an 
active area, the HPD current starts building up in response to incident light until 
beam is completely in the active area, resulting in saturation of the HPD current.
\\  

\begin{figure}[hbtp]\centering
 \includegraphics{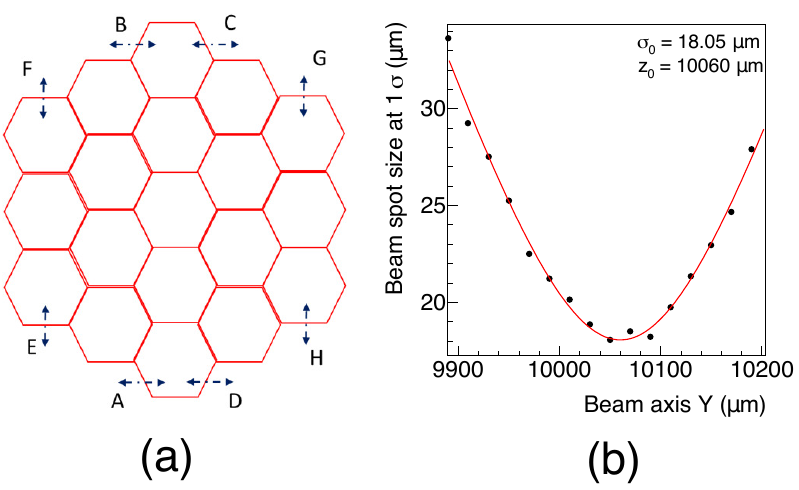}
 \caption{a) Schematic of the HPD, indicating the eight different knife-edge 
          measurement positions (dotted lines) and b) Beam spot size, obtained 
          for an HO-HPD using a 650 nm laser excitation at location C  as a 
          function of the beam axis position.}
 \label{fig:knife-edge}
\end{figure}

This method was used to characterise the laser beam profile on the photocathode 
and obtain focal plane position at eight different locations around the edges of the 
HPD as shown in Fig.~\ref{fig:knife-edge}(a). A laser beam of wavelength 650 nm 
was used for this purpose. Using the analysis procedure described in 
Ref.~\onlinecite{RShukla}, size of laser beam spot was obtained as a function of 
position of the HPD along the beam axis (Fig.~\ref{fig:knife-edge}(b)). The 
dependence of the beam spot size on position w.r.t. focal plane is fitted with a 
functional form~\cite{kogelnik1965} given in Eqn.~\ref{eqn:gaussaian_optics}. 
The minima of beam spot ($\sigma_0$) indicates the profile of the beam on 
photocathode and position of the focal plane ($z_0$). 

\begin{equation}\label{eqn:gaussaian_optics}
\sigma(z-z_0) =  \sigma_0 \times \sqrt{1+\left(\frac{{\rm M}^2 \lambda(z-z_0)}{4 \pi 
\sigma_0^{2}}\right)^2} 
\end{equation}

The size of the beam is observed to be 18.1 $\mu$m at the focal plane. 
Figure~\ref{fig:knife-edge_CCD} shows image of the HPD surface taken with the 
built-in CMOS camera. The image shows that, the fiber optic plate has a  
feature size of about 5 $\mu m$. The differential intensity distribution is observed 
to be modulated around focal plane, since the size of the light beam becomes 
comparable to the feature size of FOP, thus,  enhancing the estimation of 
measured beam width. However, note that the step size used in the 2D fine 
scan of HPD is much coarser than the measured  size of the beam spot. Knife 
edge datasets were taken at eight different positions, as shown in 
Fig.~\ref{fig:knife-edge}(a). Larger sampling across the entire HPD area was 
useful in establishing the uniformity of the focal plane position, {\it i.e.} to 
determine if any tilt is present in the HPD surface due to imperfections in the 
mechanical mounting.  Table~\ref{tab:knife-edge_summary} shows the focal 
plane positions recorded for a representative HO-HPD and HE-HPD.  A small 
variation observed in the focal plane position is due to  the small tilt present in 
the surface due to mounting imperfections. 
This tilt was then compensated dynamically by adjusting (instead of fixed) 
focal plane position at regular intervals during the 2-D scan with a weighted 
average of known focal plane positions obtained earlier with knife-edge 
method. \\

\begin{table}[]
\centering
\caption{Focal plane positions at different knife-edge scan positions on the HPD.}
\label{tab:knife-edge_summary}
\begin{tabular}{cccc}
\hline
\multirow{2}{*}{Scan Point} & \multicolumn{2}{l}
{\begin{tabular}[c]{@{}l@{}} Focal Plane Position (mm)\end{tabular}} \\
& \begin{tabular}[c]{@{}l@{}}HO HPD \end{tabular} 
& \begin{tabular}[c]{@{}l@{}}HE HPD \end{tabular} \\
\hline
A & 10.170  & 10.430 \\
E & 10.160  & 10.440 \\
B & 10.060  & 10.350 \\
F & 10.060  & 10.370 \\
C & 10.070  & 10.250 \\
G & 9.950   & 10.240 \\
D & 10.080  & 10.320 \\
H & 10.060  & 10.300 \\
\hline
\end{tabular}
\end{table}

\section{2-D Scans}\label{sec:2d-scan}

\begin{figure}[hbtp]\centering
 \includegraphics{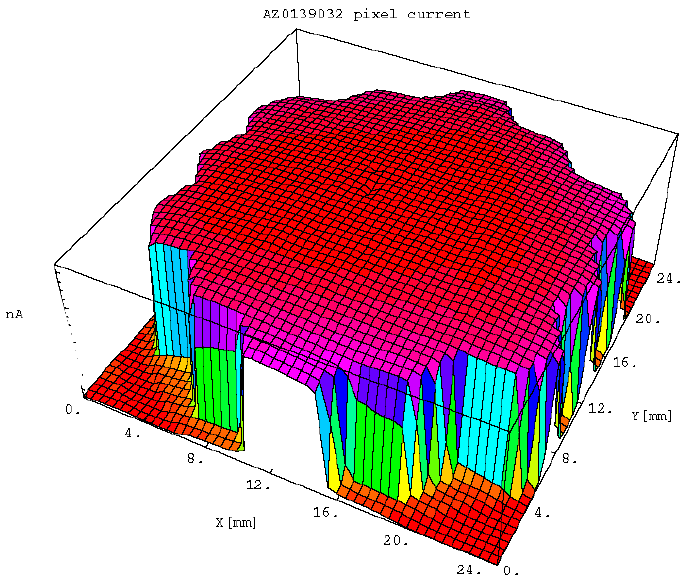}
 \caption{2D scan of an HPD carried out before its installation in the CMS 
          hadron calorimeter. Figure is adapted from Ref. \onlinecite{Cushman:2000iv}.}
 \label{fig:hpd_old_scan}
\end{figure}

Before installing the HPDs in the CMS calorimeter, coarse 2D scans of the HPD 
response were done using the setup described in 
Ref.~\onlinecite{Cushman:1103002}, which provided a laser beam of green light 
(520 nm) with beam spot size of 0.5 mm. The HPD could be moved transversely 
in steps of 0.5 mm. As can be seen from Fig.~\ref{fig:hpd_old_scan}, the response 
of the HPD was observed to be quite uniform across the entire surface of 
photocathode~\cite{Cushman:2000iv}. During entire Phase-0 period of the 
CMS-HCAL  operation, there were no HPD failures, though their performance 
gradually reduced with time and increasing integrated luminosity.  We expect 
degradation in performance to increase with increasing pseudorapidity.  However, 
at same eta, but different phi, the degradation of the signals were observed to be 
quite different. Since radiation damage is effectively independent of phi, the 
observed effect can be possibly attributed to the damage to HPDs. The HO-HPD 
was exposed to scintillation and calibration light at reduced high voltage (-6 kV) up 
to 2012 corresponding to an integrated luminosity of 6.1 $fb^{-1}$ and 23.3 $fb^{-1}$ 
at center of mass energy of 7 TeV and 8 TeV respectively. Similarly, the HE-HPD 
was exposed up to 2016 corresponding to an integrated luminosity of 6.1 $fb^{-1}$, 
23.3 $fb^{-1}$ and 45.0 $fb^{-1}$ at center of mass energy of 7, 8 and 13 TeV 
respectively. \\

Now we present results on the microscopic characterisation of an HPD after its 
decommissioning from the detector. Once the focal plane positions were determined, 
a 2-D scan of the HPD was performed after aligning the HPD surface with the focal 
plane. 2-D scans were performed with a step size ranging from 200 to 300 $\mu m$ 
to cover entire HPD area (25 mm $\times$ 25 mm), followed by finer scans with step 
size up to 75 $\mu m$ to cover specific pixels on HPD. The HPD was raster scanned 
(row by row) to record the response at each predefined step to map the response.
The focal axis position was dynamically adjusted every 
5 mm, using the weighted average of known focal positions at different points.  At 
each scan position (after setting up appropriate BV and HV), using laser illumination, 
the current flowing through the HPD was recorded as an average of 32 consecutive 
measurements. Dark current was recorded before starting each row by switching off 
the laser. Dark current measurement at the beginning of every row was necessary 
since the dark current was seen to have long term drift.  First, an HO HPD was 
scanned with 650 nm laser excitation followed by 520 nm excitations. An HE HPD 
was scanned only with 520 nm laser. Since the spectral response of the light from 
the scintillator peaks at wavelengths in the green region, green laser light was used, 
to make the extracted information relevant.

\subsection{HO-HPD Scans}

\begin{figure}[hbtp]\centering
  \includegraphics{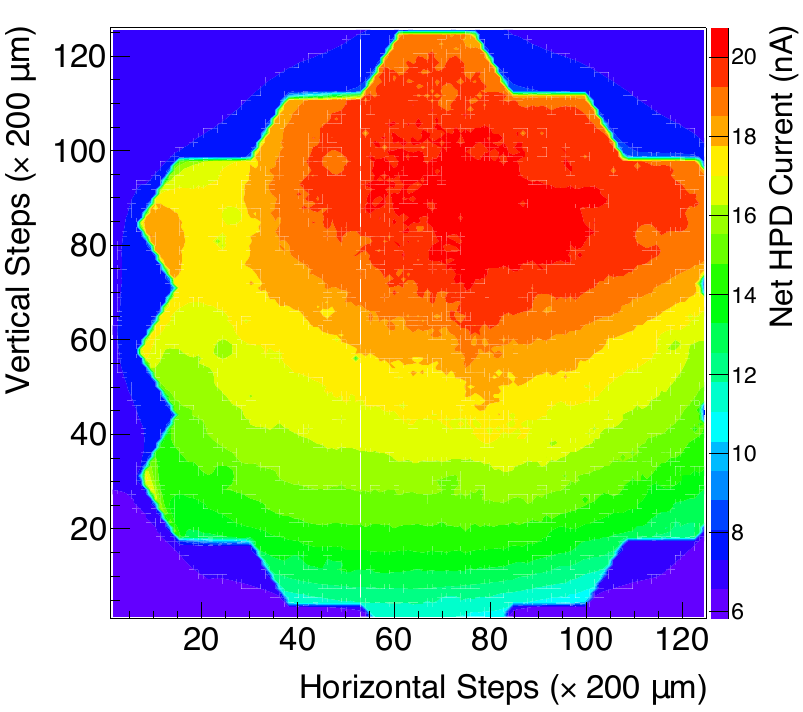}
  \caption{Net HPD current from an HO-HPD scan recorded with 650 nm laser 
                excitation and 200 $\mu m$ step size.}
  \label{fig:2-d_HO_650}
\end{figure}

\begin{figure*}[htbp]\centering
 \includegraphics{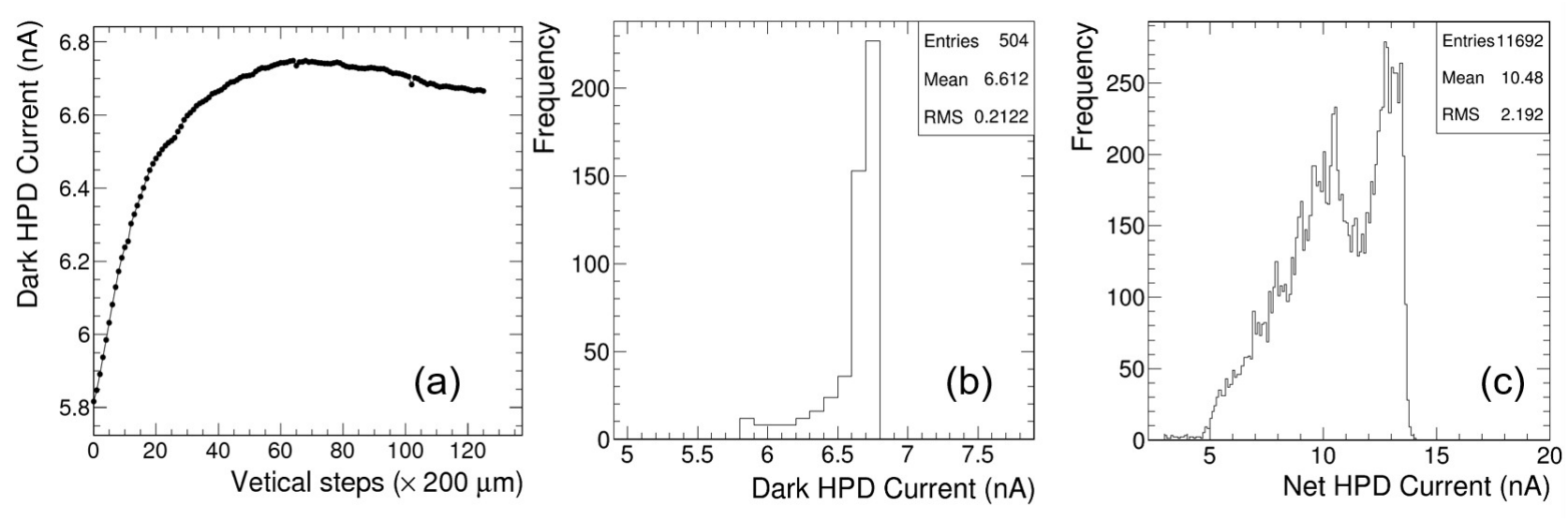}
 \caption{
   a) Dark current of HPD recorded at the beginning of each scan row. 
   b) Distribution of HPD dark current recorded at the beginning of each scan row. 
   c) Distribution of net HPD current recorded at each scan position }
 \label{fig:IHist_HO_650}
\end{figure*}

\begin{figure}[htbp]\centering
 \includegraphics{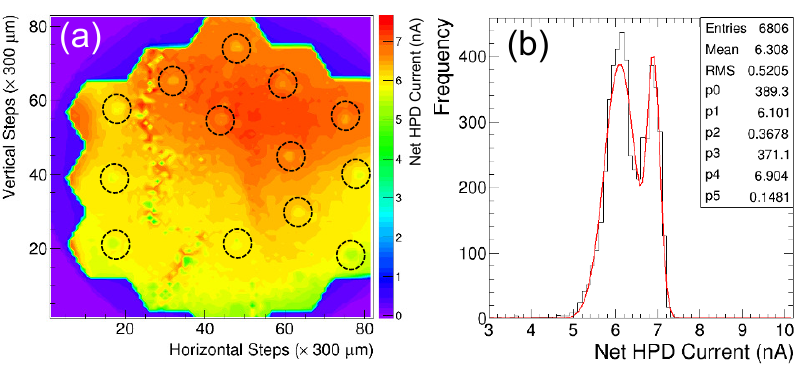}
 \caption{
    An HO-HPD 2-d scan recorded with 520 nm laser excitation and 300 $\mu m$ 
   step size: 
   a) 2-D scan of the net HPD current and 
   b) Distribution of net HPD current recorded at each scan position. }
 \label{fig:2-D_HO_520}
\end{figure}

Figure~\ref{fig:2-d_HO_650} shows the net HPD current from an HO-HPD scan 
recorded with 650 nm laser excitation with 200 $\mu m$ step size in the transverse 
direction.  The scan clearly shows variation in the net HPD current across its active 
area. The upper right region recorded a somewhat higher current than the lower 
region. The dark current  recorded at the beginning of each row was used to 
calculate the net HPD current for that row. Figure~\ref{fig:IHist_HO_650}(a) shows 
the gradual variation of dark current as a function of row number. The dark current is 
seen to increase at the beginning of the scan. The histogram of the same is shown 
in Fig.~\ref{fig:IHist_HO_650}(b). The variation in dark current during the entire run 
(RMS/Mean) is observed to be around 3\%. Figure~\ref{fig:IHist_HO_650}(c) shows 
distribution of the net HPD current in the active region. It indicates response 
variation of about 20\% (RMS/Mean). Since the  quantum efficiency  of an HPD at 
$\lambda=$ 650 nm is much lower than that for green light, the response of an 
HPD measured using $\lambda=$ 650 nm excitation is not expected to be uniform.  
The 2-D scan plots also reveal  faint degraded spots at certain localised regions in 
almost all HPD pixels. The observed localised degradation spots bring to light an 
interesting feature; hence, it was needed to be studied in more detail. To study 
these features carefully, 2-D scan of HO-HPD was performed with 520 nm laser.\\

\begin{figure*}[hbtp]\centering
 \includegraphics{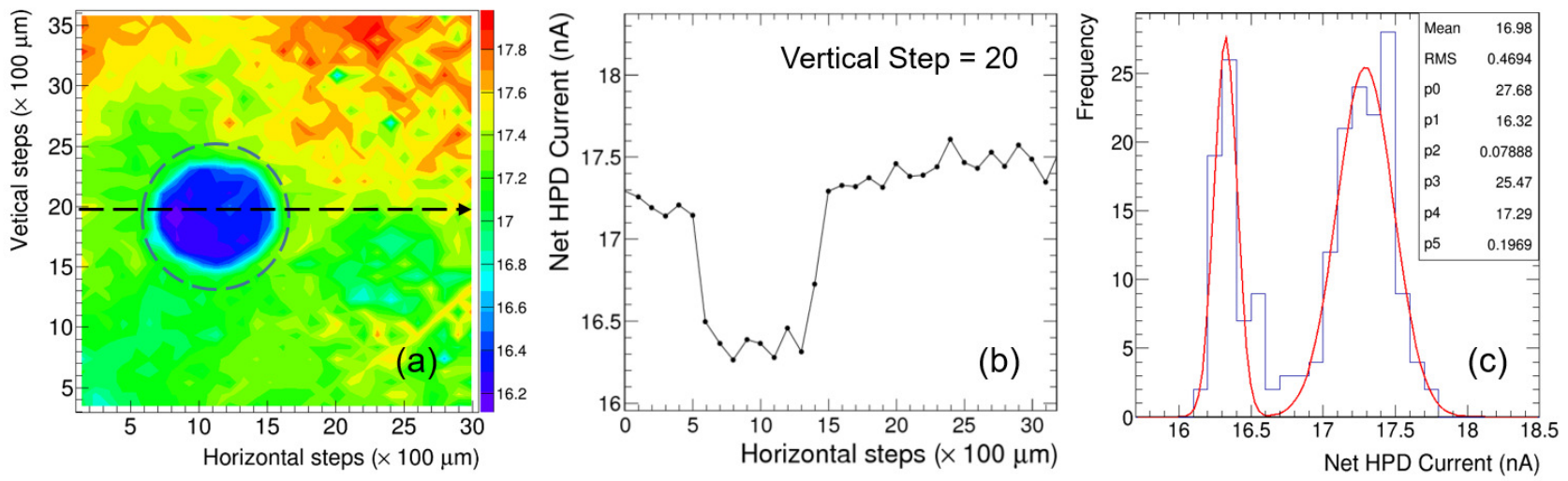}
 \caption{
   a) Localised  2D fine scan of HO-HPD recorded around fiber imprint area 
      with 520 nm laser excitation and 100 $\mu m$ step size, 
   b) Row scan passing through center of imprint (Row Number: 20) and 
   c) Distribution of net HPD current in vicinity of the fiber spot shown in a) by 
      dashed circle.}
 \label{fig:2-D_local_HO_520}
\end{figure*}

Figure~\ref{fig:2-D_HO_520}  shows the results of  2-D scans carried out with 
520 nm laser excitation. As can be seen from  
Fig.~\ref{fig:2-D_HO_520}(a), the response of photocathode to this 
wavelength is quite uniform, as compared to  that observed with the 650 nm laser. 
The histogram of net current in the active region 
(Fig.~\ref{fig:2-D_HO_520}(b)) shows a variation of about 8\%. Also, faint spots 
seen in the 650 nm data are even more prominent in this scan due to greater 
sensitivity of HPD at this wavelength ($\lambda=$ 520 nm). The localised 
degradation seems very prominent and visible in almost every pixel. This can be 
due to the calibration fiber installed in each pixel. Calibration runs are taken 
periodically by passing pulsed LED light through this fiber to monitor the gain of 
the HPD. Excessive illumination of localised regions of the photocathode 
exposed to the calibration fiber (compared to fibers coming from HO scintillators)  
may have led to comparatively higher degradation of photocathode under the 
calibration fiber area. These spots were studied in more detail with another fine 
scan taken around the imprint of calibration fiber  with a smaller step size of 
100 $\mu m$. The resulting net HPD current from the 2-D scan is shown in 
Fig.~\ref{fig:2-D_local_HO_520}(a) while Fig.~\ref{fig:2-D_local_HO_520}(b) 
shows a projected row scan of the same at one of the rows, passing roughly 
through the center of the calibration fiber imprint. A clear shadow of the size, same 
as that of the fiber, is seen on the photocathode. 
Figure~\ref{fig:2-D_local_HO_520}(c) shows a 
histogram of the net current in a region around the shadow of fiber shown in 
Fig.~\ref{fig:2-D_local_HO_520}(a). There are two clear gaussian distributions;
the left distribution corresponds to scan points under the shadow of the calibration 
fiber and the right distribution corresponds to scan points outside the shadow  
of calibration fiber. As can be seen from this histogram, a degradation in the 
photon detection efficiency of about 5\% (p1/p4) is 
observed in the photocathode area exposed to the calibration fiber w.r.t. 
neighbouring area. Though the reduction is not very significant, the overall 
setup (MROS), and the methodology adopted, demonstrates excellent 
sensitivity to study such important features of HPD response.  Subsequently, 
similar studies were also carried out for HE HPD which have gone through 
substantial radiation exposure as compared to HO-HPD. These studies 
on the HE-HPD are discussed in the following section. 

\begin{figure}[htb]\centering
 \includegraphics{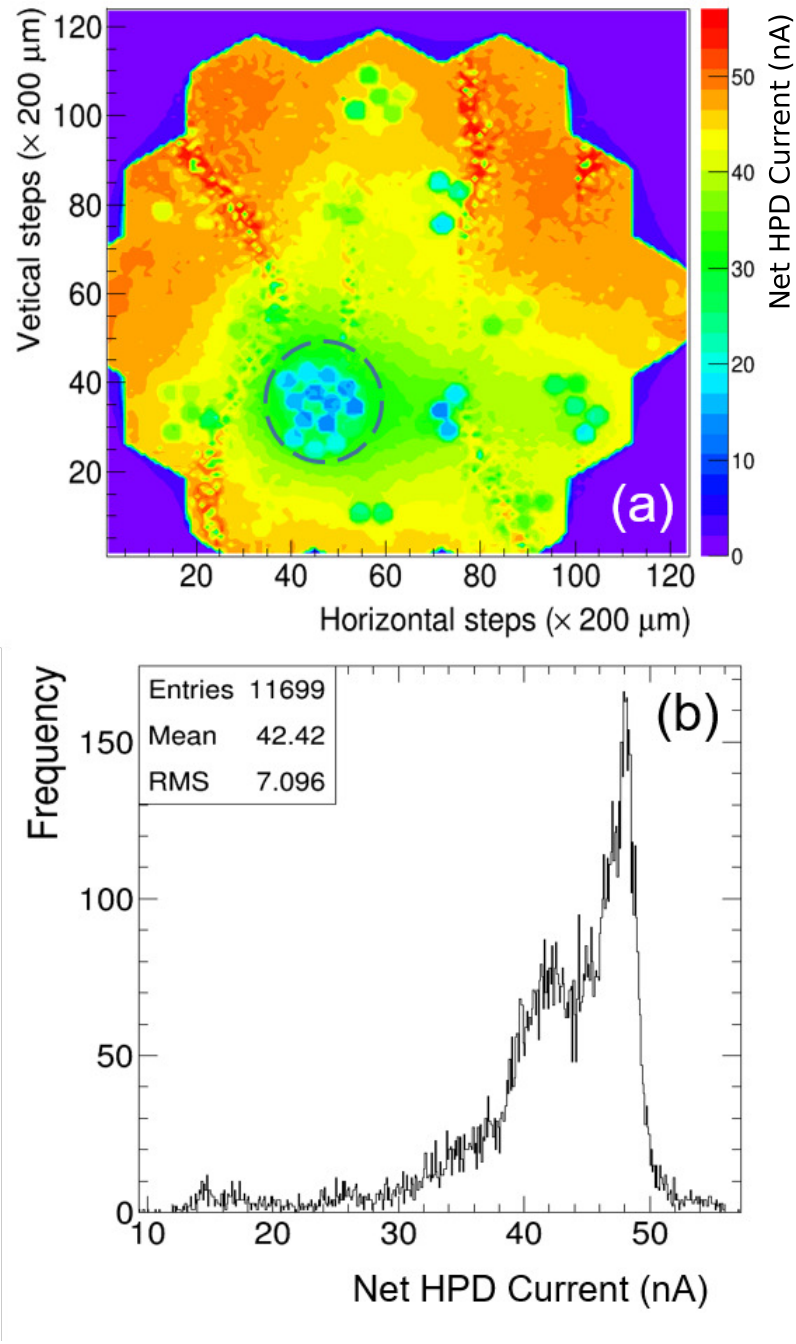}
 \caption{
   a) HE-HPD 2D scan recorded with 520 nm laser excitation and 
      200 $\mu m$ step size. 
   b) Distribution of net HPD current recorded at each scan position. }
 \label{fig:2-D_HE_520}
\end{figure}

\subsection{HE-HPD Scans}
Similar to the HO HPD scans presented in the previous section, an HPD 
decommissioned from the HE detector (HEP17 RM4) was also scanned with a 
520 nm laser beam.  A 2-D scan of the entire photocathode area of an HE-HPD 
was recorded with a step size of 200 $\mu m$. The 2D scan of net HPD current 
at each scan position is shown in Fig.~\ref{fig:2-D_HE_520}(a). The scan clearly 
shows many distinct features including several fiber imprints in almost every pixel 
of the HPD. Figure~\ref{fig:2-D_HE_520}(b) shows a histogram of the net HPD 
current, indicating overall variation (RMS/Mean) of the HPD response across 
entire surface of photocathode to be around 17\%. This is significantly larger than 
that observed in the HO-HPD at the same wavelength. As expected, this suggests 
higher overall damage of the photocathode of an HPD, extracted from the HE detector. 
Further,  the HE-HPD scan shows the presence of many circular spots of same size 
as that of the fibers coming from the scintillators. They are due to  degradation of 
photocathode under the shadow of these fibers. This can be attributed to the higher 
amount of scintillation light produced by the HE scintillators due to the large 
radiation they receive. The resulting light, which is incident on the photocathode of 
the HPD, significantly damaged the region of photocathode which is under the 
fibers coming from scintillators. It is to be noted that the photocathode of the 
HO-HPD (Fig.~\ref{fig:2-D_HO_520})  under 
the shadow of fibers coming from its scintillator did not show any degradation. 
In the HE detector, each pixel on the HPD is illuminated with a  different quantity 
of scintillation light, depending on the location and depth of the scintillator layer in
a tower. As explained above, towers with higher pseudorapidity, larger area 
of scintillator tile, and shallower depth are expected to produce larger scintillation 
light (see Fig.~\ref{fig:HPD1}(c)). Hence, pixels mapping to such towers should 
see a higher degradation of the photocathode. \\

\begin{figure*}[hbtp]\centering
 \includegraphics{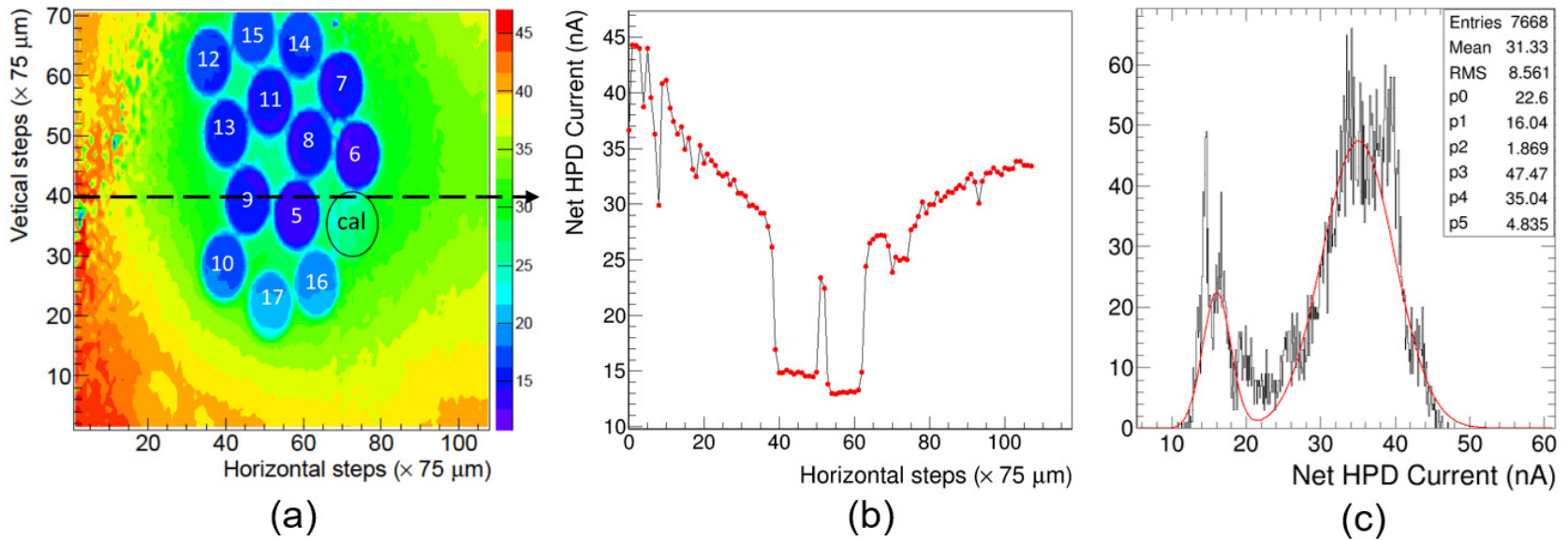}
 \caption{a) 2D fine scan of pixel 14 of HE-HPD with 520 nm laser and 
             75 $\mu m$ step size. Number within each fiber spot represents the layer 
             number of that fiber. 
          b) Sectional view of the 2D fine scan data for pixel 14 of the HE HPD. 
          c) Distribution of net HPD current recorded at each scan position.}
  \label{fig:2-D_local_HE_520}
\end{figure*}

A pixel with a higher number of visible fiber imprints (pixel 14, tower 
28R containing 13 layers of scintillators starting from layer number 5), 
shown by dashed circle in Fig.~\ref{fig:2-D_HE_520}(a) was 
investigated further with another finer scan of the area under this pixel, recorded 
with step size of 75 $\mu$m. Tower 28R is closer to the beam 
pipe, and hence it is expected to have higher damage. As can be seen 
from Fig.~\ref{fig:2-D_local_HE_520}(a), almost all fiber imprints (corresponding 
to scintillator fibers) in this pixel are clearly visible. They possess different 
shades, indicating varying damage, depending on the depth of the scintillator 
layers in this tower (see Fig.~\ref{fig:HPD1}(c)). The degradation due to 
calibration fiber is not clearly visible for this HPD. 
Figure~\ref{fig:2-D_local_HE_520}(b) shows a sectional plot (taken near 
row 40), which clearly shows that the photocathode area under the fibers from 
scintillators has been damaged significantly. The extent of the average damage 
for this pixel was estimated by making a histogram of the net current recorded 
at each scan point shown in  Fig.~\ref{fig:2-D_local_HE_520}(c). The histogram 
has two distinct distributions 
fitted with a Gaussian functions to extract the mean and sigma for each 
distribution. The left distribution corresponds to the part of the photocathode 
area covered by the fibers, and the right distribution shows rest of the area, 
with no fiber cover. The ratio of peaks (p1/p4) of these two distributions shows 
a relative reduction in the photocathode 
efficiency of $\sim 46\%$. It is to be noted that this reduction is only for that 
area of photocathode that is exposed to fibers coming from scintillators.
Variation of the response within the fiber imprint area is observed to be 
about 10\% (p2/p1). \\

\begin{figure}[htbp]\centering
 \includegraphics{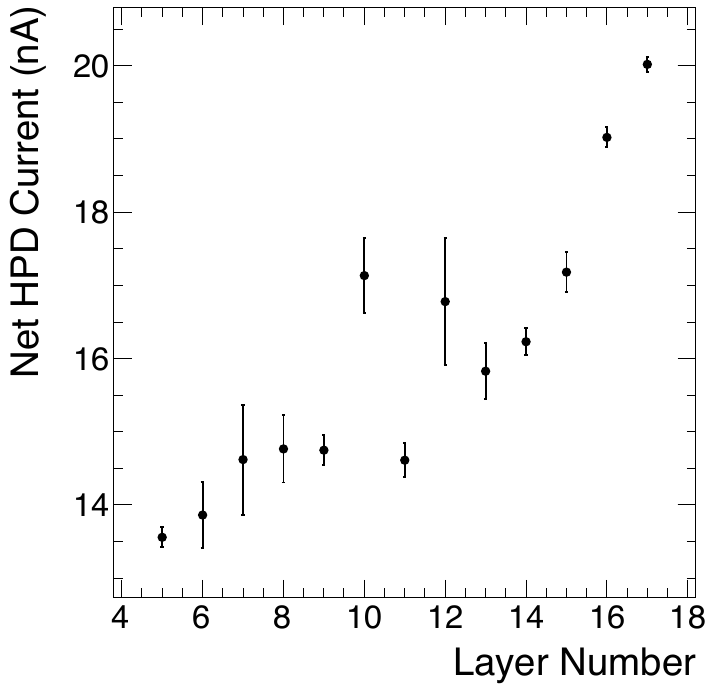}
 \caption{Plot of net average HPD current for each fiber as a function of its 
          layer number for Pixel 14 of the HE-HPD. }
 \label{fig:pix14_layer_response}
\end{figure}

Due to high resolution of the scan, the same data for this pixel  can also be 
used to deduce the  layer dependence of damage for a given tower on 
HPD pixel. Scintillators at lower depth (smaller layer number) should cause 
more damage compared to those further away from the interaction region 
(higher layer number). Tower 28R, read by pixel 14, has 13 fibers reading 
scintillation light from layer numbers 5 to 17. The layer mapping of each 
fiber spot was done using the real optical decoder unit (ODU) used for 
this HPD (ODU 19.4.27 from RM4-HEP 17). Mapping of layer number to 
corresponding fiber spot is shown in Fig.~\ref{fig:2-D_local_HE_520}(a).
The average photo response of each layer was obtained by sampling 
those points on a row scan (passing through the center of fiber) that are 
within the fiber under consideration. The response of  the photocathode 
under each fiber of each layer as a function of layer number is shown in 
Fig.~\ref{fig:pix14_layer_response}.  A  strong correlation between the 
relative photocathode response and corresponding layer number can be 
seen from this figure. As expected, the farthest  layer (L17) shows least 
damage and the innermost layer (L5) shows highest damage. The 
response of L17  is observed to be $\sim 46\%$ higher than L5. However, 
in order to have good jet energy resolution, all the layers should  have 
uniform response within $\pm$5\%. Hence, such large variation in the 
HPD response across different layers of the same tower affects the 
energy resolution of the jets produced in pp collisions that impact the HE 
calorimeter. \\

\begin{figure*}[htbp]\centering
 \includegraphics{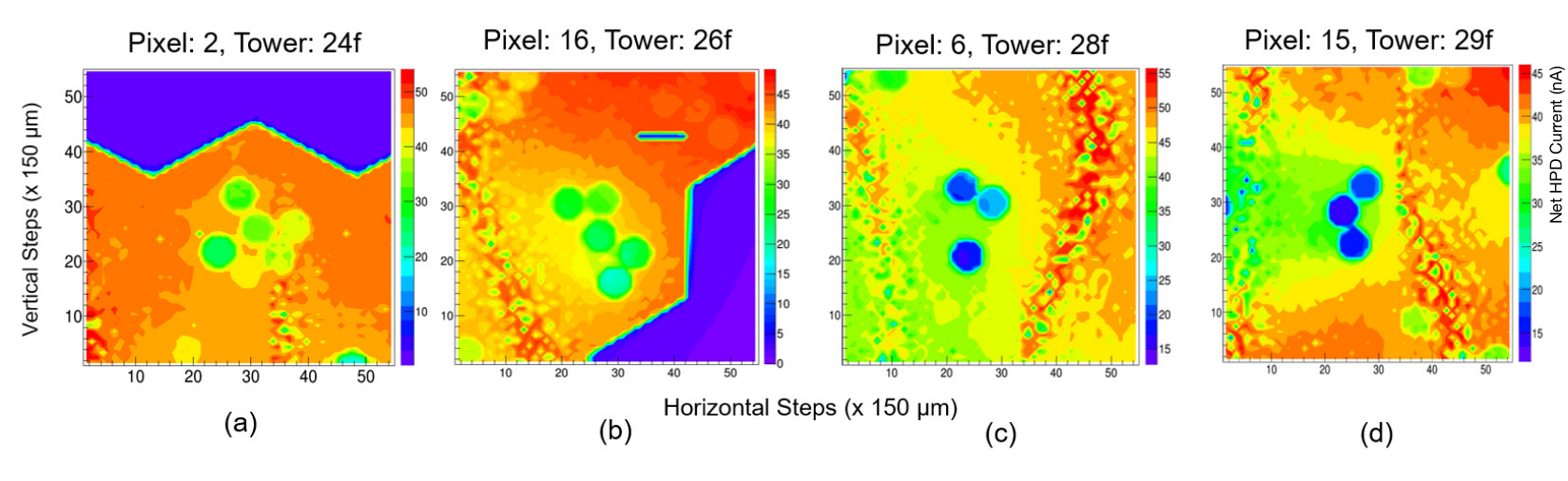}
 \caption{2D fine scans of the HE HPD recorded at four different pixels with 
          laser of 520 nm excitation and step size of 150 $\mu m$}
 \label{fig:HE_fineScans_520}
\end{figure*}

A few more fine scans, with a step size of 150 $\mu$m, were recorded 
around the damaged area for four other pixels mapping to different 
pseudorapidity towers in the HE detector. The location of the tower for 
each scanned pixel  is obtained using Fig.~\ref{fig:HPD1}(b) and 
Fig.~\ref{fig:HPD1}(c). The 2-D scan of net HPD current  for pixels 
2, 6, 15 and 16 are shown in Fig.~\ref{fig:HE_fineScans_520}(a) - (d). 
The overall degradation of each pixel is obtained using the same procedure 
described above. Table~\ref{tab:HE_fine_scan_eta} shows the degradation 
observed in five pixels in a single HPD that map to different $\eta$ towers. 
The measured degradation of the photocathode corresponding to each pixel 
confirms that the damage is larger at higher pseudorapidity regions due to 
higher scintillation light produced by high radiation levels, thus resulting in 
higher erosion of photocathode.  \\

\begin{table}[]
\centering
\caption{Summary of the local damage observed in different HE HPD 
         pixels. Tower index is followed by depth (f: front and r: rear)}
\label{tab:HE_fine_scan_eta}
\begin{tabular}{lllll}
\hline
Pixel &   No. of   & Tower   &  pseudorapidity   & Damage    \\
No.   &   layers   & Index   &                   &    (\%)   \\
\hline
   2   &    6   &   24f   &  2.17   &  25      \\
 16   &    6   &   26f   &  2.50   &  41      \\
 14   &  13   &   28r   &  2.87   &  54      \\
   6   &    3   &   28f   &  2.87   &  55      \\
 15   &    3   &   29f   &  3.00   &  59      \\
\hline
\end{tabular}
\end{table}

\section{Breakdown Voltage Uniformity}\label{sec:breakdown}
As discussed in the previous sections, several 2-D scans of  HE-HPDs showed 
significant non-uniformity in the response of the  photocathode. Degradation of the
photocathode is caused by excessive illumination by scintillation light. 
Localised areas on the photocathode directly exposed to the scintillation light 
through fibers have been damaged severely, whereas other part of the
photocathode, which is not exposed to the scintillation light, do not indicate any 
visible damage. This observation is expected to manifest as a possible 
variation in the breakdown voltage of the HPD, as well as a change in the
I-V response, measured at damaged and normal locations on 
photocathode. To ascertain this, I-V characteristic of the HE-HPD was 
measured at all 13 damaged spots on pixel 14 (tower 28R) as well as at
normal positions on the photocathode using the following procedure: 
a) HPD with all pixels  was biased with a constant BV of 80 Volts, 
b) predefined location was illuminated with a focused laser beam 
of wavelength 520 nm using the MROS setup and c) with these 
conditions, high voltage (HV) was ramped up from 0 to -7000 Volts in 
steps of 250 Volts and at each step, the HPD current 
was recorded. The dark current was noted by switching off the laser 
at the beginning of ramp up and end of ramp down cycle at each position.\\

\begin{figure}[htbp]\centering
   \includegraphics{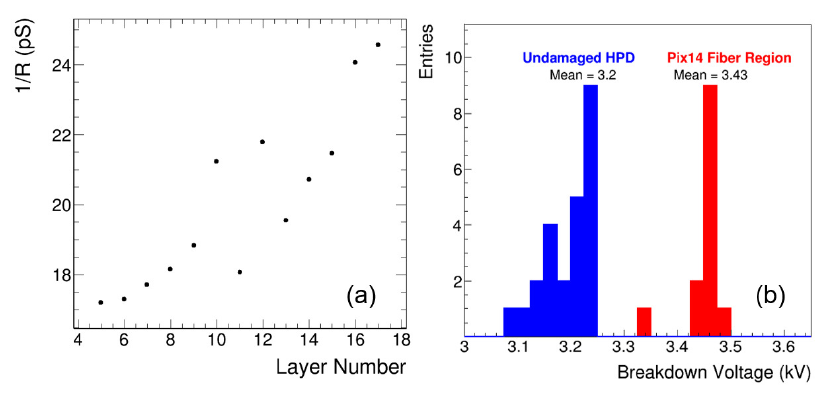}
   \caption{a) Conductance as a function of layer number  and 
            b) Distribution  of breakdown voltage for a normal  (blue) 
               and a damaged region (red) of the photocathode obtained 
               using I-V characteristics recorded at different locations.}
  \label{fig:HE_IV_dist_global_local}
\end{figure}

Centre position of each fiber imprint on the pixel-14 of the photocathode 
was identified using the 2D scans taken earlier. Similarly, several positions 
which did not show any damage were also identified using the same data. 
At each of these pre-defined positions, the I-V data was taken using the 
procedure described above. Typical I-V characteristics is shown in 
Fig.~\ref{fig:HPD-HV-IV}. The breakdown voltage is obtained by taking the 
intersection of the lines obtained by fitting the data in two different regions 
(below and above the breakdown region) as shown in the same figure. The 
slope of second line (above the breakdown region) 
represents the conductance of photodiode (dI/dV) at that position. The 
conductance depends on the flux of incident photoelectrons emitted by the
photocathode. For the same intensity of laser, this flux is expected to be lower 
for lower layer number and thus should have smaller conductance. As can be 
seen from Fig.~\ref {fig:HE_IV_dist_global_local}(a), the conductance, indeed, 
strongly depends on the layer number and follows the same trend as observed 
in 2D-scan of the HPD (Fig.~\ref{fig:pix14_layer_response}). Thus, analysis of 
I-V data provides an independent verification of a localised damage of the
photocathode that is proportional to the incident scintillation light. 
Figure~\ref {fig:HE_IV_dist_global_local}(b) shows the distribution of breakdown 
voltage measured at normal and damaged locations. The breakdown voltage 
at damaged locations is observed to be  about 200 Volts higher compared 
to that at locations with a normal  photocathode. The increase in the breakdown 
voltage is expected due to localised thickening of the dead layer on the silicon 
surface. \\

\section{Summary and Conclusions}
Independent studies of damage to HPDs used in the CMS hadron calorimeter due
to excess exposure to scintillation light were important to assess its impact on the 
detection and measurement of the energies of jets produced in the proton-proton 
collisions. The Micron Resolution Optical Scanner (MROS), built for microscopic 
characterisation of photodetectors was leveraged for this purpose. It was suitably 
modified to house HPDs, as well as its associated data acquisition system. 
Microscopic characterisation of two HPDs, decommissioned from the HO and the 
HE detector, was carried out using MROS with  two types of focused laser beams 
of  650 nm and 520 nm wavelengths. Before characterising the HPDs, the profile 
of the laser beam on the surface of the photocathode was carried out to establish 
the size of the beam spot. It was measured to be 18.1 $\mu$m, much finer than the 
step size (75-300 $\mu$m) used for performing the HPD scan. Subsequently, 
several 2-D scans were carried out with different step sizes; lower step sizes were 
used to study smaller regions of interest. The results of these studies presented in 
this article clearly demonstrate degradation of the photocathode of the HPDs from 
both detectors. The overall variation of the photocathode response was observed 
to be  8\% and 17\% for the HO and the HE HPD, respectively.  \\

Moreover, several circular features in the 2D scans of net HPD current were seen for 
HPDs from both detectors. The diameters of these spots were the same as that of 
fibers carrying either scintillation light or calibration light. This clearly indicates 
localised damage of the photocathode in the areas directly exposed to the light 
coming through fibers. Though the HPD from the HO detector did not show any 
damage due to scintillation light, a significant localised damage of photocathode 
caused by excessive illumination due to scintillation light was observed in the HPD 
decommissioned from the HE detector. It is to be noted that the HO detector is 
located behind the HB detector; thus the fluence of charged particles 
through the HO scintillators is significantly smaller than that through the HE 
scintillators. As expected,  larger damage of the photocathode was seen for those 
fibers collecting scintillation light from scintillators located at higher pseudorapidity 
and shallower depth. This region has much higher radiation generated by 
proton-proton collisions. Fine scans recorded around one such damaged pixel 
(Pixel 14, connected to tower 28R with 13 layers) show that the photocathode 
damage is higher for fibers originating from shallower depth scintillators (closer to 
the proton-proton collision point). In addition, the part of the photocathode exposed 
to fibers from the highest layer number registered $\sim 46\%$ higher net HPD 
current compared to that observed with lowest depth. In the beginning of CMS 
physics run, all these layers in a given tower were shown to have uniform  
response within $\pm 5$\%.  A differential degradation of $\sim 46\%$ can 
significantly deteriorate the jet energy resolution of the calorimeter.  Similar fine 
scans for several pixels mapping different  pseudorapidities were done for the 
HE-HPD. As expected, the  damage was observed to be proportional to the  
pseudorapidity. \\

Non-uniform response of the photocathode of the HE-HPD, observed using fine photo 
scans with focused laser light, implies that the breakdown voltage and conductance
measured at damaged and normal regions of photocathode need not be the same.  
Measurement revealed that, indeed, the breakdown voltage measured at the damaged 
region ($\sim$ 3.4 kV) is about  200 Volts higher than that measured in the normal 
region of photocathode. Also, the conductance was observed to be higher for higher 
layer numbers. The dependence of conductance on layer number 
(Fig.~\ref{fig:HE_IV_dist_global_local}(a)) is quite similar to that observed in the 2D 
scans (Fig.~\ref {fig:pix14_layer_response}). Hence, both the methods, that are 
independent of each other, confirm localised damage of photocathode proportional 
to the incident scintillation light.

\begin{acknowledgments}

We thank CMS-HCAL members for their contribution to this effort.  We are grateful 
to Vasken Hagopian and Sarah Eno 
for their valuable feedback that has helped in improving the manuscript. We thank 
S. Chavan for his help in making mechanical fixtures and setting up the 
system. We thank Ravindra Verma for his help in taking the data. \\
\end{acknowledgments}

\bibliography{DN-18-004-v5}

\providecommand{\noopsort}[1]{}\providecommand{\singleletter}[1]{#1}%
\begin{thebibliography}{17}%
\makeatletter
\providecommand \@ifxundefined [1]{%
 \@ifx{#1\undefined}
}%
\providecommand \@ifnum [1]{%
 \ifnum #1\expandafter \@firstoftwo
 \else \expandafter \@secondoftwo
 \fi
}%
\providecommand \@ifx [1]{%
 \ifx #1\expandafter \@firstoftwo
 \else \expandafter \@secondoftwo
 \fi
}%
\providecommand \natexlab [1]{#1}%
\providecommand \enquote  [1]{``#1''}%
\providecommand \bibnamefont  [1]{#1}%
\providecommand \bibfnamefont [1]{#1}%
\providecommand \citenamefont [1]{#1}%
\providecommand \href@noop [0]{\@secondoftwo}%
\providecommand \href [0]{\begingroup \@sanitize@url \@href}%
\providecommand \@href[1]{\@@startlink{#1}\@@href}%
\providecommand \@@href[1]{\endgroup#1\@@endlink}%
\providecommand \@sanitize@url [0]{\catcode `\\12\catcode `\$12\catcode
  `\&12\catcode `\#12\catcode `\^12\catcode `\_12\catcode `\%12\relax}%
\providecommand \@@startlink[1]{}%
\providecommand \@@endlink[0]{}%
\providecommand \url  [0]{\begingroup\@sanitize@url \@url }%
\providecommand \@url [1]{\endgroup\@href {#1}{\urlprefix }}%
\providecommand \urlprefix  [0]{URL }%
\providecommand \Eprint [0]{\href }%
\providecommand \doibase [0]{http://dx.doi.org/}%
\providecommand \selectlanguage [0]{\@gobble}%
\providecommand \bibinfo  [0]{\@secondoftwo}%
\providecommand \bibfield  [0]{\@secondoftwo}%
\providecommand \translation [1]{[#1]}%
\providecommand \BibitemOpen [0]{}%
\providecommand \bibitemStop [0]{}%
\providecommand \bibitemNoStop [0]{.\EOS\space}%
\providecommand \EOS [0]{\spacefactor3000\relax}%
\providecommand \BibitemShut  [1]{\csname bibitem#1\endcsname}%
\let\auto@bib@innerbib\@empty
\bibitem [{CMS(1997)}]{CMS:1997tfa}%
  \BibitemOpen
  \href {http://cdsweb.cern.ch/record/357153} {\enquote {\bibinfo {title}
  {{CMS: The hadron calorimeter technical design report}},}\ }\bibinfo {type}
  {Tech. Rep.}\ (\bibinfo {year} {1997})\BibitemShut {NoStop}%
\bibitem [{\citenamefont {Baiatian}\ \emph {et~al.}(2007)\citenamefont
  {Baiatian}, \citenamefont {Sirunyan}, \citenamefont {Emeliantchik},
  \citenamefont {Massolov} \emph {et~al.}}]{Baiatian:1049915}%
  \BibitemOpen
  \bibfield  {author} {\bibinfo {author} {\bibfnamefont {G.}~\bibnamefont
  {Baiatian}}, \bibinfo {author} {\bibfnamefont {A.~M.}\ \bibnamefont
  {Sirunyan}}, \bibinfo {author} {\bibfnamefont {I.}~\bibnamefont
  {Emeliantchik}}, \bibinfo {author} {\bibfnamefont {V.}~\bibnamefont
  {Massolov}},  \emph {et~al.} (\bibinfo {collaboration} {CMS HCAL
  Collaboration}),\ }\href {https://cds.cern.ch/record/1049915} {\enquote
  {\bibinfo {title} {{Design, Performance, and Calibration of CMS Hadron-Barrel
  Calorimeter Wedges}},}\ }\bibinfo {type} {Tech. Rep.}\ \bibinfo {number}
  {CMS-NOTE-2006-138}\ (\bibinfo  {institution} {CERN},\ \bibinfo {address}
  {Geneva},\ \bibinfo {year} {2007})\BibitemShut {NoStop}%
\bibitem [{\citenamefont {Abdullin}\ \emph
  {et~al.}(2008{\natexlab{a}})\citenamefont {Abdullin}, \citenamefont
  {Abramov}, \citenamefont {Acharya} \emph {et~al.}}]{Abdullin2008}%
  \BibitemOpen
  \bibfield  {author} {\bibinfo {author} {\bibfnamefont {S.}~\bibnamefont
  {Abdullin}}, \bibinfo {author} {\bibfnamefont {V.}~\bibnamefont {Abramov}},
  \bibinfo {author} {\bibfnamefont {B.}~\bibnamefont {Acharya}},  \emph
  {et~al.},\ }\href {\doibase 10.1140/epjc/s10052-008-0573-y} {\bibfield
  {journal} {\bibinfo  {journal} {The European Physical Journal C}\ }\textbf
  {\bibinfo {volume} {55}},\ \bibinfo {pages} {159} (\bibinfo {year}
  {2008}{\natexlab{a}})}\BibitemShut {NoStop}%
\bibitem [{\citenamefont {Baiatian}\ \emph {et~al.}(2008)\citenamefont
  {Baiatian}, \citenamefont {Sirunyan}, \citenamefont {Emeliantchik} \emph
  {et~al.}}]{Baiatian:1103003}%
  \BibitemOpen
  \bibfield  {author} {\bibinfo {author} {\bibfnamefont {G.}~\bibnamefont
  {Baiatian}}, \bibinfo {author} {\bibfnamefont {A.~M.}\ \bibnamefont
  {Sirunyan}}, \bibinfo {author} {\bibfnamefont {I.}~\bibnamefont
  {Emeliantchik}},  \emph {et~al.} (\bibinfo {collaboration} {CMS HCAL
  Collaboration}),\ }\href {https://cds.cern.ch/record/1103003} {\enquote
  {\bibinfo {title} {{Design, Performance, and Calibration of CMS Hadron Endcap
  Calorimeters}},}\ }\bibinfo {type} {Tech. Rep.}\ \bibinfo {number}
  {CMS-NOTE-2008-010}\ (\bibinfo  {institution} {CERN},\ \bibinfo {address}
  {Geneva},\ \bibinfo {year} {2008})\BibitemShut {NoStop}%
\bibitem [{\citenamefont {Abdullin}\ \emph
  {et~al.}(2008{\natexlab{b}})\citenamefont {Abdullin}, \citenamefont
  {Abramov}, \citenamefont {Acharya} \emph {et~al.}}]{Abdullin2008HO}%
  \BibitemOpen
  \bibfield  {author} {\bibinfo {author} {\bibfnamefont {S.}~\bibnamefont
  {Abdullin}}, \bibinfo {author} {\bibfnamefont {V.}~\bibnamefont {Abramov}},
  \bibinfo {author} {\bibfnamefont {B.}~\bibnamefont {Acharya}},  \emph
  {et~al.},\ }\href {\doibase 10.1140/epjc/s10052-008-0756-6} {\bibfield
  {journal} {\bibinfo  {journal} {The European Physical Journal C}\ }\textbf
  {\bibinfo {volume} {57}},\ \bibinfo {pages} {653} (\bibinfo {year}
  {2008}{\natexlab{b}})}\BibitemShut {NoStop}%
\bibitem [{\citenamefont {Acharya}\ \emph {et~al.}(2006)\citenamefont
  {Acharya}, \citenamefont {Aziz}, \citenamefont {Banerjee} \emph
  {et~al.}}]{Acharya:973131}%
  \BibitemOpen
  \bibfield  {author} {\bibinfo {author} {\bibfnamefont {B.~S.}\ \bibnamefont
  {Acharya}}, \bibinfo {author} {\bibfnamefont {T.}~\bibnamefont {Aziz}},
  \bibinfo {author} {\bibfnamefont {S.}~\bibnamefont {Banerjee}},  \emph
  {et~al.},\ }\href {https://cds.cern.ch/record/973131} {\enquote {\bibinfo
  {title} {{The CMS Outer Hadron Calorimeter}},}\ }\bibinfo {type} {Tech.
  Rep.}\ \bibinfo {number} {CMS-NOTE-2006-127}\ (\bibinfo  {institution}
  {CERN},\ \bibinfo {address} {Geneva},\ \bibinfo {year} {2006})\BibitemShut
  {NoStop}%
\bibitem [{\citenamefont {Abdullin}\ \emph
  {et~al.}(2008{\natexlab{c}})\citenamefont {Abdullin}, \citenamefont
  {Abramov}, \citenamefont {Acharya} \emph {et~al.}}]{Abdullin2008B}%
  \BibitemOpen
  \bibfield  {author} {\bibinfo {author} {\bibfnamefont {S.}~\bibnamefont
  {Abdullin}}, \bibinfo {author} {\bibfnamefont {V.}~\bibnamefont {Abramov}},
  \bibinfo {author} {\bibfnamefont {B.}~\bibnamefont {Acharya}},  \emph
  {et~al.},\ }\href {\doibase 10.1140/epjc/s10052-007-0459-4} {\bibfield
  {journal} {\bibinfo  {journal} {The European Physical Journal C}\ }\textbf
  {\bibinfo {volume} {53}},\ \bibinfo {pages} {139} (\bibinfo {year}
  {2008}{\natexlab{c}})}\BibitemShut {NoStop}%
\bibitem [{\citenamefont {Kryshkin}\ and\ \citenamefont
  {Ronzhin}(1986)}]{KRYSHKIN1986583}%
  \BibitemOpen
  \bibfield  {author} {\bibinfo {author} {\bibfnamefont {V.}~\bibnamefont
  {Kryshkin}}\ and\ \bibinfo {author} {\bibfnamefont {A.}~\bibnamefont
  {Ronzhin}},\ }\href {\doibase https://doi.org/10.1016/0168-9002(86)90420-1}
  {\bibfield  {journal} {\bibinfo  {journal} {Nuclear Instruments and Methods
  in Physics Research Section A: Accelerators, Spectrometers, Detectors and
  Associated Equipment}\ }\textbf {\bibinfo {volume} {247}},\ \bibinfo {pages}
  {583 } (\bibinfo {year} {1986})}\BibitemShut {NoStop}%
\bibitem [{Note1()}]{Note1}%
  \BibitemOpen
  \bibinfo {note} {DEP (Delft Electronic Products), Dwazziewegen 2\\ Roden,9301
  Netherlands}\BibitemShut {NoStop}%
\bibitem [{\citenamefont {Elias}(1997)}]{ELIAS1997104}%
  \BibitemOpen
  \bibfield  {author} {\bibinfo {author} {\bibfnamefont {J.}~\bibnamefont
  {Elias}},\ }\href {\doibase https://doi.org/10.1016/S0168-9002(96)00971-0}
  {\bibfield  {journal} {\bibinfo  {journal} {Nuclear Instruments and Methods
  in Physics Research Section A: Accelerators, Spectrometers, Detectors and
  Associated Equipment}\ }\textbf {\bibinfo {volume} {387}},\ \bibinfo {pages}
  {104 } (\bibinfo {year} {1997})},\ \bibinfo {note} {new Developments in
  Photodetection}\BibitemShut {NoStop}%
\bibitem [{\citenamefont {Cushman}, \citenamefont {Heering},\ and\
  \citenamefont {Ronzhin}(2000)}]{Cushman:2000iv}%
  \BibitemOpen
  \bibfield  {author} {\bibinfo {author} {\bibfnamefont {P.}~\bibnamefont
  {Cushman}}, \bibinfo {author} {\bibfnamefont {A.}~\bibnamefont {Heering}}, \
  and\ \bibinfo {author} {\bibfnamefont {A.}~\bibnamefont {Ronzhin}},\
  }\bibfield  {booktitle} {\emph {\bibinfo {booktitle} {{BEAUNE 1999 New
  Developments in Photodetection: Proceedings of the 2nd International
  Conference on New Developments in Photodetection (NDIP99), Beaune, France,
  June 21-25, 1999}}},\ }\href {\doibase 10.1016/S0168-9002(99)01236-X}
  {\bibfield  {journal} {\bibinfo  {journal} {Nucl. Instrum. Meth.}\ }\textbf
  {\bibinfo {volume} {A442}},\ \bibinfo {pages} {289} (\bibinfo {year}
  {2000})}\BibitemShut {NoStop}%
\bibitem [{\citenamefont {Cushman}\ \emph {et~al.}(1997)\citenamefont
  {Cushman}, \citenamefont {Heering}, \citenamefont {Nelson}, \citenamefont
  {Timmermans}, \citenamefont {Dugad}, \citenamefont {Katta},\ and\
  \citenamefont {Tonwar}}]{CUSHMAN1997107}%
  \BibitemOpen
  \bibfield  {author} {\bibinfo {author} {\bibfnamefont {P.}~\bibnamefont
  {Cushman}}, \bibinfo {author} {\bibfnamefont {A.}~\bibnamefont {Heering}},
  \bibinfo {author} {\bibfnamefont {J.}~\bibnamefont {Nelson}}, \bibinfo
  {author} {\bibfnamefont {C.}~\bibnamefont {Timmermans}}, \bibinfo {author}
  {\bibfnamefont {S.}~\bibnamefont {Dugad}}, \bibinfo {author} {\bibfnamefont
  {S.}~\bibnamefont {Katta}}, \ and\ \bibinfo {author} {\bibfnamefont
  {S.}~\bibnamefont {Tonwar}},\ }\href {\doibase
  https://doi.org/10.1016/S0168-9002(96)00972-2} {\bibfield  {journal}
  {\bibinfo  {journal} {Nuclear Instruments and Methods in Physics Research
  Section A: Accelerators, Spectrometers, Detectors and Associated Equipment}\
  }\textbf {\bibinfo {volume} {387}},\ \bibinfo {pages} {107 } (\bibinfo {year}
  {1997})},\ \bibinfo {note} {new Developments in Photodetection}\BibitemShut
  {NoStop}%
\bibitem [{\citenamefont {Cushman}, \citenamefont {Heering},\ and\
  \citenamefont {Ronzhin}(1998)}]{CUSHMAN1998300}%
  \BibitemOpen
  \bibfield  {author} {\bibinfo {author} {\bibfnamefont {P.}~\bibnamefont
  {Cushman}}, \bibinfo {author} {\bibfnamefont {A.}~\bibnamefont {Heering}}, \
  and\ \bibinfo {author} {\bibfnamefont {A.}~\bibnamefont {Ronzhin}},\ }\href
  {\doibase https://doi.org/10.1016/S0168-9002(98)00691-3} {\bibfield
  {journal} {\bibinfo  {journal} {Nuclear Instruments and Methods in Physics
  Research Section A: Accelerators, Spectrometers, Detectors and Associated
  Equipment}\ }\textbf {\bibinfo {volume} {418}},\ \bibinfo {pages} {300 }
  (\bibinfo {year} {1998})}\BibitemShut {NoStop}%
\bibitem [{\citenamefont {Cushman}\ and\ \citenamefont
  {Sherwood}(2007)}]{Cushman:1103002}%
  \BibitemOpen
  \bibfield  {author} {\bibinfo {author} {\bibfnamefont {P.}~\bibnamefont
  {Cushman}}\ and\ \bibinfo {author} {\bibfnamefont {B.}~\bibnamefont
  {Sherwood}},\ }\href {https://cds.cern.ch/record/1103002} {\enquote {\bibinfo
  {title} {{Lifetime Studies of the 19-channel Hybrid Photodiode for the CMS
  Hadronic Calorimeter}},}\ }\bibinfo {type} {Tech. Rep.}\ \bibinfo {number}
  {CMS-NOTE-2008-011}\ (\bibinfo  {institution} {CERN},\ \bibinfo {address}
  {Geneva},\ \bibinfo {year} {2007})\BibitemShut {NoStop}%
\bibitem [{\citenamefont {Shukla}\ \emph {et~al.}(2014)\citenamefont {Shukla},
  \citenamefont {Dugad}, \citenamefont {Garde}, \citenamefont {Gopal},
  \citenamefont {Gupta},\ and\ \citenamefont {Prabhu}}]{RShukla}%
  \BibitemOpen
  \bibfield  {author} {\bibinfo {author} {\bibfnamefont {R.}~\bibnamefont
  {Shukla}}, \bibinfo {author} {\bibfnamefont {S.}~\bibnamefont {Dugad}},
  \bibinfo {author} {\bibfnamefont {C.}~\bibnamefont {Garde}}, \bibinfo
  {author} {\bibfnamefont {A.}~\bibnamefont {Gopal}}, \bibinfo {author}
  {\bibfnamefont {S.}~\bibnamefont {Gupta}}, \ and\ \bibinfo {author}
  {\bibfnamefont {S.}~\bibnamefont {Prabhu}},\ }\href {\doibase
  10.1063/1.4863880} {\bibfield  {journal} {\bibinfo  {journal} {Review of
  Scientific Instruments}\ }\textbf {\bibinfo {volume} {85}},\ \bibinfo {pages}
  {023301} (\bibinfo {year} {2014})}\BibitemShut {NoStop}%
\bibitem [{\citenamefont {Firester}, \citenamefont {Heller},\ and\
  \citenamefont {Sheng}(1977)}]{firester1977}%
  \BibitemOpen
  \bibfield  {author} {\bibinfo {author} {\bibfnamefont {A.~H.}\ \bibnamefont
  {Firester}}, \bibinfo {author} {\bibfnamefont {M.~E.}\ \bibnamefont
  {Heller}}, \ and\ \bibinfo {author} {\bibfnamefont {P.}~\bibnamefont
  {Sheng}},\ }\href@noop {} {\bibfield  {journal} {\bibinfo  {journal} {APPLIED
  OPTICS}\ }\textbf {\bibinfo {volume} {16}} (\bibinfo {year}
  {1977})}\BibitemShut {NoStop}%
\bibitem [{\citenamefont {Kogelnik}(1965)}]{kogelnik1965}%
  \BibitemOpen
  \bibfield  {author} {\bibinfo {author} {\bibfnamefont {H.}~\bibnamefont
  {Kogelnik}},\ }\href@noop {} {\bibfield  {journal} {\bibinfo  {journal}
  {APPLIED OPTICS}\ }\textbf {\bibinfo {volume} {4}} (\bibinfo {year}
  {1965})}\BibitemShut {NoStop}%
\end{thebibliography}%

\end{document}